\documentstyle [twocolumn,prl,aps,floats,graphicx] {revtex}

\begin{document}
\draft
\flushbottom
\begin{title}
{\bf Multiply charged metal cluster anions}
\end{title} 
\author{Constantine Yannouleas,$^1$ Uzi Landman,$^1$ Alexander Herlert,$^2$
and Lutz Schweikhard$^2$} 
\address{
$^1$School of Physics, Georgia Institute of Technology,
Atlanta, Georgia 30332-0430\\
$^2$Institut f\"{u}r Physik, Johannes-Gutenberg Universit\"{a}t, D-55099 Mainz, 
Germany}
\date{November 2000}
\maketitle

\begin{abstract}
Formation, stability patterns, and decay channels of silver dianionic and gold 
trianionic clusters are investigated with Penning-trap experiments and 
a shell-correction method including shape deformations. The theoretical
predictions pertaining to the appearance sizes and electronic shell effects 
are in remarkable agreement with the experiments. Decay of the multiply
anionic clusters occurs predominantly by electron tunneling through a
Coulomb barrier, rather than via fission, leading to appearance sizes 
unrelated to those of multiply cationic clusters. 
\end{abstract}

\narrowtext

Charging of macroscopic metal spheres is an old subject with scientific
accounts dating back to Coulomb and Faraday \cite{fara}. 
Recently, electrical charging emerged as a central issue in
connection with quantal nanostructures in diverse areas of condensed-matter, 
molecular, and cluster physics, such as artificially fabricated semiconductor 
nanodevices known as Quantum Dots \cite{kast1}, and gas-phase microsystems, 
such as fullerenes \cite{hl,yl1}, large organic molecules \cite{wang}, 
and metal microclusters \cite{bms,bl,yl2,yl3,kasperovich}. 

In this Letter, we investigate the physical processes underlying
generation of multiply {\it anionic\/} gas-phase metal clusters,
$M_N^{Z-}$. The production and observation of gas-phase multiply anionic 
aggregates had remained for many years a challenging experimental goal.
This state of affairs changed in the last few years
with the observation of several {\it dianionic\/} aggregates, including 
doubly negative fullerenes \cite{hl}, 
metal-tetrahalide molecular dianions \cite{kapp}, and most recently dianionic 
gold clusters \cite{schw3}. A few further observations of dianionic 
\cite{schw1} and {\it trianionic\/} \cite{schw2,stoermer} gas-phase metal 
clusters have also been reported. In addition, we note that a few theoretical 
studies \cite{yl4} of multiply charged anionic metal clusters have also 
appeared and that the earlier of them [16(a)] had anticipated 
several of the recent experimental developments;
however, overall, the field of multiply anionic aggregates remains at
its early stages.

Unlike the case of macroscopic metal spheres, where the number of 
positive or negative elementary charges that can be added is essentially
arbitrary, with
gas-phase metal clusters this number is size-controlled due to 
quantum-confinement instabilities arising from the restriction of the excess 
charges to a very small space. In this Letter, we provide definitive evidence
that multiply anionic metal clusters (irrespective of $Z$) behave in a way
remarkably different from their cationic counterparts. That is,
they dissociate via a different decay channel, i.e., electron autodetachment 
through a repulsive Coulomb barrier (CB) \cite{note1},
instead of fragmenting via fission. 
This results in appearance sizes unrelated to 
those known for the cationic species \cite{bms,yl2}. 
In addition, we show that the multiply
charged metal-cluster anions exhibit electronic shell effects 
arising from magic major shells and subshells associated with ellipsoidal
shapes \cite{yl2,yl3}, in full analogy with their neutral and cationic 
counterparts.
We arrive at these conclusions through a careful and systematic comparison of 
theoretical predictions and original experimental observations for both doubly
and triply anionic metal clusters. In particular, we investigate 
the stability and decay channels of Ag$_N^{2-}$ and Au$_N^{3-}$, and determine
the corresponding appearance sizes $n^{Z-}_a$, where clusters with 
$N < n^{Z-}_a$ are energetically unstable.  

Theoretically, two classes of decay channels need to be considered
for the appearance sizes of the M$_N^{Z-}$ clusters:\\ 
(I) Binary (for $Z=2$ and $Z=3$) and ternary 
(for $Z=3$) fission channels,
\begin{equation}
M^{2-}_N \rightarrow M^-_P + M^-_{N-P}~,
\label{eq1}
\end{equation}
\begin{equation}
M^{3-}_N \rightarrow M^{2-}_P + M^-_{N-P}~,
\label{eq2}
\end{equation}
\begin{equation}
M^{3-}_N \rightarrow M^-_P + M^-_Q + M^-_{N-P-Q}~,
\label{eq3}
\end{equation}
which have well known analogs in the case of multiply cationic clusters
\cite{bms,bl,yl2} and atomic nuclei \cite{book}, and\\
(II) Electron autodetachment via tunneling through a repulsive
Coulombic barrier \cite{yl4}, 
\begin{equation}
M^{Z-}_N \rightarrow M^{(Z-1)-}_N + e~;\;\; Z \geq 2~,
\label{eq4}
\end{equation}
in analogy to proton and alpha decay in atomic nuclei \cite{book,naza}.
For analysis of the energetics of these channels, we use
a finite-temperature semi-empirical
shell-correction method (SCM), which incorporates triaxial
shapes and which has been previously used successfully to describe 
the properties of neutral and cationic metal clusters \cite{yl2,yl3}.
The finite-temperature multiple electron affinities of a cluster of $N$
atoms of valence $v$ (we take $v=1$ for Au and Ag) are defined as
\[
A_Z(N,\beta)=F(\beta, vN, vN+Z-1)-F(\beta,vN,vN+Z)~,
\]
where $F$ is the free energy, $\beta=1/k_B T$, and $Z \geq 1$ is
the number of excess electrons in the cluster (e.g., the first, second, and
third electron affinities correspond to $Z=1$, $Z=2$, and $Z=3$ respectively).
In the SCM, the free energy $F$ is separated into 
a smooth liquid-drop-model (LDM) part $\widetilde{F}_{\text{LDM}}$ (varying 
monotonically with $N$), and a Strutinsky-type shell-correction term 
$\Delta F_{\text{sp}}=
F_{\text{sp}}-\widetilde{F}_{\text{sp}}$. $F_{\text{sp}}$ is the 
canonical (fixed $N$ at a given $T$) free energy of the valence electrons.
The latter are
treated as independent single particles moving in an effective mean-field 
potential (approximated by a modified Nilsson hamiltonian pertaining to 
triaxial cluster shapes).
$\widetilde{F}_{\text{sp}}$ is the Strutinsky-averaged free energy.
The smooth $\widetilde{F}_{\text{LDM}}$ contains volume, surface, and
curvature contributions, whose coefficients and temperature dependencies 
are determined as described in Ref.\ \cite{yl3}.
In addition to the finite-temperature contribution due to the 
electronic entropy, the entropic contribution from thermal shape fluctuations 
is evaluated via a Boltzmann averaging [9(b)].  

The smooth contribution $\widetilde{A}_Z(N,\beta)$ to the 
full multiple electron affinities ${A}_Z(N,\beta)$ can be approximated 
by the LDM expression [16(a)],  
\begin{equation}
\widetilde{A}_Z = \widetilde{A}_1 -\frac{(Z-1)e^2}{R(N)+\delta_0}
= W -\frac{(Z-1+\gamma)e^2}{R(N)+\delta_0}~,
\label{eq6}
\end{equation}
where $R(N)=r_s N^{1/3}$ is the radius of the positive background ($r_s$ is
the Wigner-Seitz radius which depends weakly on $T$ due to volume dilation),
$\gamma=5/8$, $\delta_0$ is an electron spillout parameter, and the work 
function $W$ is assumed to be temperature independent [we take $W$(Au)$=5.31$ 
eV and $W$(Ag)$=4.26$ eV].

\begin{figure}[t]
\centering\includegraphics[width=7.0cm]{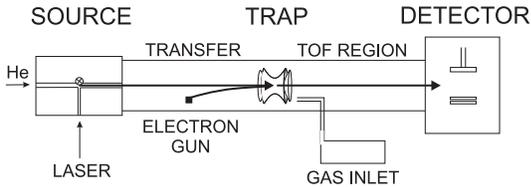}\\
~~~\\
\caption{
Overview of the experimental setup.
}
\end{figure}
The present experimental results have been achieved by use of a Penning
trap system devoted to metal cluster research (Fig.\,1) \cite{exsetup}: 
A Smalley-type ion source produces the
clusters by laser vaporization of a metal wire in the presence of a
helium gas pulse \cite{clsource}. 
The resulting clusters
are either neutral or singly charged. The anionic species are transferred
by ion-optical elements through differential-pumping stages to a Penning
trap where they are captured in flight \cite{schnatz}.
The  storage is provided by the
combination of a 5T homogeneous magnetic field and an electric trapping
potential along the magnetic field lines \cite{brown}. In order
to generate higher charge states, an electron beam of about 40 eV is sent
through the trap. Simultaneously supplied argon atoms are ionized
producing low-energy secondary electrons which stay stored in the
trap volume. Some of these electrons attach to
the clusters, thus charging them up. This reaction is analyzed by
ejection of the clusters from the trap and time-of-flight mass
spectrometry. Single-ion counting has to be performed, since only a few
tens of ions are observed in each experimental cycle. The signal
intensity of several hundred sequences is summed to increase the 
statistical significance of the data.

\begin{figure}[t]
\centering\includegraphics[width=7.0cm]{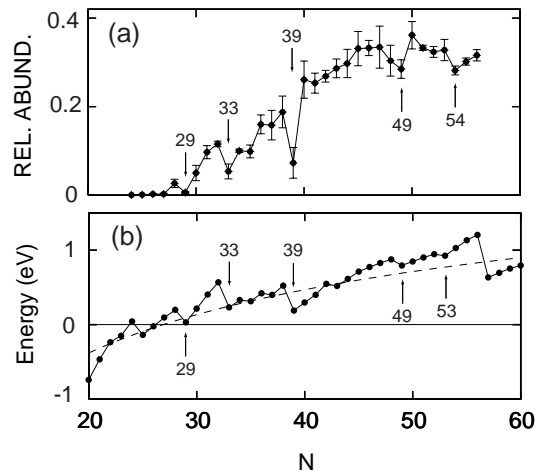}\\
~~~\\
\caption{ 
(a) Experimental yields of dianionic silver clusters Ag$_N^{2-}$,
plotted versus cluster size. Most measurements have been repeated (some 
several times). The error bars indicate the statistical uncertainty or the 
deviations between different measurements (whatever is larger).
(b) Theoretical SCM second electron affinities $A_2$ for Ag$_N^{2-}$
clusters at $T=300$ K. LDM results [see
Eq.\ (\ref{eq6}) with $Z=2$] are depicted by the dashed line.
}
\end{figure}
The conversion yield from mono- to di- and trianionic metal clusters is
a complex function of many experimental parameters. The details of the
process still await further elucidation. Nevertheless, when care is
taken that the parameters are kept unchanged during an experimental
sequence, valuable information on the clusters' properties can be gained
from the relative conversion yields. In particular, based on the case of gold 
cluster dianions, it has been suggested [13,16(b)]  
that the resulting experimental yield pattern
can be correlated to the clusters' multiple electron affinities.
The cases of Ag$_N^{2-}$ and Au$_N^{3-}$ considered here show this to be
a general property applicable to any excess charge and cluster species.

Fig.\ 2(a) displays the observed relative abundances of the dianionic
silver clusters. Dianionic Ag$_N^{2-}$ have been observed for  $N \geq 24$.
In general, the yield is very low for
$N<28$ and then increases linearly
up to about size $N = 45$ where it levels off. Embedded in this general
trend are pronounced dips at $N = 29, 33$, and 39, as well as around 
$N = 49$ and 54. 

To gain further insight into these experimental trends,  
Fig.\,2(b) presents the theoretical SCM results for 
the second electron affinity $A_2$ of silver clusters 
in the size range $20 \leq N \leq 60$. $N=24$ with 
$A_2=0.04$ eV forms a very weak
island in front of the main stability branch starting at $N=27$. 
Disregarding this weak island, the appearance size for the main stability
island of Ag$_N^{2-}$ (i.e., the smallest size with 
$A_2 > 0$) is predicted to be $n_a^{2-}$(Ag)$=27$. 
The theoretical curve for the electron
affinities and the experimental measurements correlate to a remarkable
degree. In particular: (I) The smallest size seen in the
experiment is $N=24$, but the yields from $N=24$ to 27 are very small;
(II) At $N=27$, the experimental yields start on the average to increase
sharply and keep increasing for $27 \leq N \leq 45$, in agreement 
with the theoretical prediction for
$n_a^{2-}$(Ag) and with the trend of the theoretical curve;
and (III) The dips in the measured yields [marked by arrows in Fig.\ 2(a)]
are in accordance with corresponding minima in the theoretical curve. 

Underlying the theoretical and experimental
patterns shown in Fig.\,2 are electronic shell effects [compare in Fig.\,2(b) 
the shell-corrected results indicated by the solid line with the LDM curve] 
combined with energy-lowering shape deformations for the open-shell
clusters \cite{yl3}. These deformations are akin to Jahn-Teller distortions 
and are associated with the lifting of spherical spectral degeneracies and the 
formation of electronic subshells (for 30, 34, and 50 electrons) 
in addition to the (magic) major shell (for 40 electrons).

\begin{figure}[t]
\centering\includegraphics[width=7.0cm]{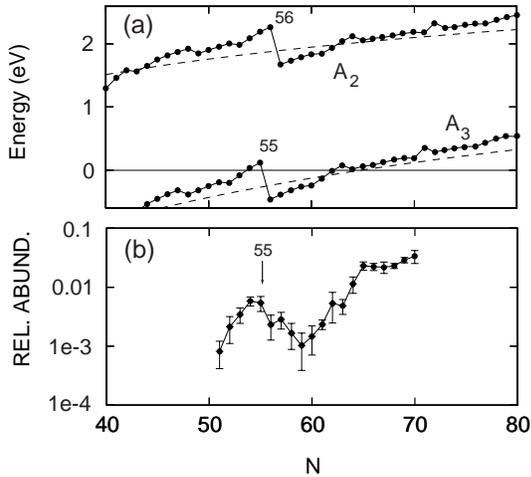}\\
~~~~\\
\caption{ 
(a) Calculated second ($A_2$, upper curve) and third ($A_3$, lower curve) 
electron affinities for Au$_N$ clusters at $T=300$ K in the size range 
$40 \leq N \leq 80$. Theoretical results
from SCM calculations are connected by a solid line, and LDM
results [see Eq.\ (\ref{eq6}) with $Z=2$ and 3] are depicted by the dashed 
lines.
(b) Experimental average relative abundances of Au$_N^{3-}$
as compared to Au$_N^{2-}$ clusters in the same size range
(error bars as in Fig.\,2). Notice the logarithmic scale.
}
\end{figure}
Next we discuss the case of trianionic metal clusters.
In Fig.\,3(a), we display the SCM theoretical results \cite{note2,note3} for 
the second, $A_2$, and third, $A_3$, electron affinities of Au$_N$ clusters in 
the size range $40 \leq N \leq 80$. Multiply anionic clusters M$_N^{Z-}$ 
with $A_Z < 0$ are unstable \cite{yl4} against electron emission via
tunneling through a CB. In the case of doubly anionic gold clusters, all sizes
in the aforementioned range are stable, i.e., they have $A_2 > 0$
[see upper curve in Fig.\ 3(a)]. In contrast, for the
gold-cluster trianions, those with $N \leq 53$ and $56 \leq N \leq 62$ have
$A_3 < 0$. Thus, the appearance size for Au$_N^{3-}$ is $n_a^{3-} \approx
54$. Apparently, the major shell closure at 58 electrons 
(corresponding to the Au$_{56}^{2-}$ parent of the triply charged 
Au$_{56}^{3-}$ cluster) creates an island of 
stability (the clusters with $N=54$ and 55 have $A_3 > 0$) preceeding the main 
stability branch (with $N \geq 63$).

These theoretical results correlate again remarkably well with the measured
relative abundance spectrum for Au$_N^{3-}$ in the same size 
range [see Fig.\,3(b)]. In particular, note the overall similarity between
the trends of $A_3$ in Fig.\,3(a) and the yields for the trianions in 
Fig.\,3(b), that is, the presence of a weak promontory around $N=55$ 
in front of the principal branch of the detected Au$_N^{3-}$ clusters. 
Comparison of the shell-corrected results (solid dots) with the LDM curve 
(dashed line) further highlights that electronic shell effects underlie the 
calculated detailed patterns for $A_3$ shown in Fig.\,3(a) and those
observed experimentally in Fig.\,3(b).

In addition, our measurements revealed that: (I) Dianionic gold clusters are 
observed, as expected, for all sizes under investigation ($50\leq N\leq 70$), 
since $n_a^{2-}$(Au)$=12$ [13,16(b)];  
and (II) The yields of the stable 
gold-cluster trianions are two orders of magnitude smaller
than the ones corresponding to the parent dianions (which are roughly 
constant), reflecting the difference in the magnitiudes of the associated 
multiple electron affinities [see theoretical curves for $A_2$ and $A_3$ in
Fig.\,3(a)].

\begin{figure}[t]
\centering\includegraphics[width=7.0cm]{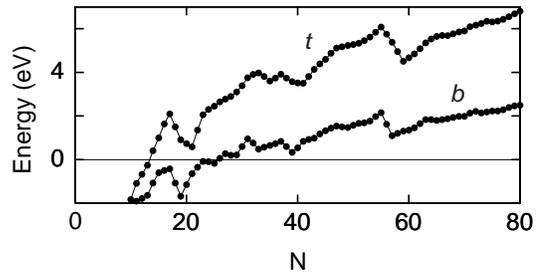}\\
~~~~\\
\caption{ 
Theoretical SCM fission dissociation energies for the most favorable channel
($\Delta^{3-}_{N,\gamma}$ in eV) at $T=300$ K for binary 
($\gamma=b$, lower curve) and ternary ($\gamma=t$,upper curve) fission
of trianionic Au$_N^{3-}$ clusters. Exothermic fission 
($\Delta^{3-}_{N,\gamma} < 0$) is found only for the smallest 
clusters with $N \leq 25$.
}
\end{figure}
To explore the energetic stability of the Au$_N^{3-}$ clusters against
binary $(b)$ and ternary $(t)$ fission [see Eqs.\ (\ref{eq2}) and 
(\ref{eq3})], we show in Fig.\,4 SCM results for the fission dissociation 
energies, $\Delta^{3-}_{N,\gamma}$ $(\gamma=b$ or $t)$
associated with the most favorable channel for a given parent 
Au$^{3-}_N$ cluster. We found that the most favorable channel 
corresponds to the generation of one or two closed-shell Au$^-$ anions
in the case of binary and ternary fission, respectively. Thus
$\Delta^{3-}_{N,b} = F$(Au$^-)+F($Au$^{2-}_{N-1})-F($Au$^{3-}_N)$ and
$\Delta^{3-}_{N,t} = 2F$(Au$^-)+F($Au$^-_{N-2})-F($Au$^{3-}_N)$,
with the total free energies of the multiply anionic parent and the charged 
fission products calculated at $T=300$ K.
The fission results summarized in Fig.\,4
for the most favorable channel illustrate that exothermic fission
(that is $\Delta^{3-}_{N,\gamma} < 0$) occurs only for the 
smallest sizes ($N \leq 25$). This, together with the existence of a fission 
barrier, leads us to conclude that the decay of Au$_N^{3-}$ clusters is 
dominated by the electron autodetachment process [see Eq.\ (\ref{eq4})], 
which is operative when $A_3 < 0$ and involves tunneling through a 
CB \cite{yl4}, rather than by fission \cite{note4}.

The significant findings about the generation, stability patterns, and
decay channels, obtained here via Penning-trap experiments and 
theoretical SCM investigations, include: (i) A remarkable agreement
between the theoretically predicted and the measured appearance sizes
[$n^{2-}_a$(Ag)$=$27 and $n^{3-}_a$(Au)$=$54]; (ii) The strong
influence of electronic shell effects on the measured yields; and
(iii) Identification of electron autodetachment, rather than fragmentation
via fission, as the prevalent decay channel (irrespective of the 
magnitude of the excess negative charge), leading to appearance sizes 
unrelated to those known from the more familiar case of multiply charged 
cationic clusters. In light of these findings, it will be of interest
to explore further the properties of stored multiply-charged
metal cluster anions by either collisional or laser activation 
\cite{krueckeberg}, as well as investigate their chemical properties
\cite{sanc}.


This research is supported by grants from the U.S. Department of Energy
(Grant No. FG05-86ER45234), the Deutsche Forschungsgemeinschaft, the
EU networks "EUROTRAPS" and "CLUSTER COOLING", the Materials Science
Research Center Mainz and the Fonds der Chemischen Industrie.


\begin{references}
\bibitem{note1}
The relevance of this decay channel for metal clusters was first suggested
in the theoretical investigations of Ref.\ [16(a)],  
where the appearance sizes of Na$_N^{Z-}$, K$_N^{Z-}$, and Al$_N^{Z-}$ 
$(2 \leq Z \leq 4)$ were estimated using the LDM expression (\ref{eq6}).
\bibitem{note2}
For the Au$_N$ and Ag$_N$ clusters discussed here ($N \leq 80$),
our results at $T=0$ K and $T=300$ K differ only slightly. 
For cases where the SCM reveals significant thermal effects 
(i.e., for higher temperatures and/or larger $r_s$'s 
as in the case of Na$_N$ and K$_N$ clusters), 
see Ref.\ [9(b)]. 
\bibitem{note3}
For Au$_N^{Z-}$(Ag$_N^{Z-}$) clusters, the $T=0$ parameters 
entering in the SCM calculation (for definitions see Ref.\ 
[9(a)]) are  
$U_0=-0.045$, $r_s=3.01$ a.u., $t=0.37$(0.47) a.u., $\delta_0=1.31$ a.u.,
$\delta_2=0$, $W=5.31$(4.26) eV, $\alpha_v=-8.06$ eV, $\alpha_s=2.52$(2.05) 
eV, and $\alpha_c=1.04$(0.86) eV. For an additional explanation, 
see footnote 17 in Ref.\ [9(b)]. 
The temperature dependence of the surface
tension and the coefficient of linear thermal expansion were taken from
standard tables.
\bibitem{note4}
A similar conclusion can be drawn for the case of
doubly anionic metal clusters (see Ref.\ [16(b)]).  
\end{references}
\end{document}